\documentclass[12pt,preprint,psfig]{aastex}

\long\def\HII{H\,{\sc ii}}
\long\def\Ha{H$\alpha$}           
\long\def\kms{~km\,s$^{-1}$}

\long\def\Lsun{~L$_{\odot}$}
\long\def\Msun{~M$_{\odot}$}
\long\def\h{$^{\rm h}$}
\long\def\m{$^{\rm m}$}
\long\def\s{$^{\rm s}$}

\long\def\ATCA{{\it ATCA}}
\long\def\Chandra{{\it Chandra}}
\long\def\proplyd{ProPlyD-like object}
\long\def\plike{ProPlyD-like}

\shorttitle{ProPlyD-like objects in NGC~3603}
\shortauthors{M\"ucke et al.}

\begin{document}
\title{ATCA radio imaging of the ProPlyD-like objects  \\
       in the giant \HII\ region NGC~3603}

\author{A.~M\"ucke\altaffilmark{1}, B.S.~Koribalski\altaffilmark{2},
        A.F.J.~Moffat\altaffilmark{1}, M.F.~Corcoran\altaffilmark{3},
        I.R.~Stevens\altaffilmark{4} }

\altaffiltext{1}{Universit\'e de Montr\'eal, D\'epartement de Physique,
                 C.P. 6128, Succ. Centre-Ville, Montr\'eal, QC,
		 H3C 3J7, Canada, and Observatoire du Mont M\'egantic; AM now at:
		 Ruhr-Universit\"at Bochum, Institut f\"ur Theoretische Physik, Lehrstuhl IV:
		 Weltraum- und Astrophysik, D-44780 Bochum, Germany\\
                 email: {\it{afm@tp4.ruhr-uni-bochum.de}},
                 {\it{moffat@astro.umontreal.ca}}}
\altaffiltext{2}{Australia Telescope National Facility, CSIRO,
                 P.O. Box 76, Epping 1710, Australia\\
                 email: {\it{Baerbel.Koribalski@atnf.csiro.au}}}
\altaffiltext{3}{Goddard Space Flight Center, NASA/GSFC Code 661,
                 Laboratory for High Energy Astrophysics, Greenbelt,
		 MD 20771, USA\\
                 email: {\it{corcoran@barnegat.gsfc.nasa.gov}}}
\altaffiltext{4}{School of Physics \& Astronomy, University of Birmingham,
                 Birmingham B15 2TT, UK\\
                 email: {\it{irs@star.sr.bham.ac.uk}}}

%
%
\begin{abstract}
Three cometary-shaped objects in the giant \HII\ region NGC~3603, originally
found and identified as proto-planetary disks (ProPlyDs) by \cite{brandner2}
using HST+VLT in the optical and near-infrared, have been detected with the
Australia Telescope Compact Array (\ATCA\thanks{The Australia Telescope is
funded by the Commonwealth of Australia for operation as a National Facility
managed by CSIRO.}) in the radio continuum at 3 and 6~cm. All three {\plike}
objects are clearly resolved with an extent of a few arcseconds. The integrated
6~cm fluxes are up to 1.3 times higher than the 3~cm fluxes with spectral 
indices averaged over the whole clump between $\alpha=-0.1$ and --0.5
($S_\nu \propto \nu^{\alpha}$), indicating the likely presence of non-thermal emission
in at least some of the sources.
We present spectral index maps, and show that the sites of negative
radio spectral indices are predominantly concentrated in the direction of the
tails in at least two of the three {\plike} nebulae while positive spectral indices are found
in the region facing the ionizing star cluster. We propose that thermal
bremsstrahlung and non-thermal synchrotron radiation are at work
in all three {\plike} sources. In at least one
of the three objects optically thin non-thermal synchrotron emission
appears to dominate when averaged over their whole spatial extent, while
the spectrum of the second source shows a marginal indication of a
non-thermal spectrum.
The average spectrum of the third source is in agreement with
thermal bremsstrahlung. All
measured fluxes are at least one order of magnitude higher than those
predicted by \cite{brandner2}. Upper limits for mass loss rates due to
photo-evaporation are calculated to be $\sim 10^{-5}$~\Msun\,year$^{-1}$ and for
electron densities to be $\sim 10^{4}$~cm$^{-3}$. Due to the unexpectedly large
radio luminosities of the {\plike} features and because the radio emission
is extended a (proto-)stellar
origin of the non-thermal emission from a dust enshrouded star appears
unlikely. Instead we propose that magnetized regions within the envelope of
the {\plike} nebulae exist.  
\end{abstract}


\section{Introduction}
The giant \HII\ region NGC~3603, located at a distance of about 6~kpc, shows
the densest concentration and largest collection of visible massive stars 
known in our Galaxy. Recent HST images of $\sim$0\farcs2 angular resolution 
\citep{moffat} have shown that NGC~3603 consists of three Wolf-Rayet (WR)
stars and $\sim$70 O-type stars, with an estimated 40--50 of these stars 
located in 
the central $\sim$30\arcsec$\times$30\arcsec\ (1 pc $\times$
1 pc) region. The three H-rich WR-stars of subtype WNL are also located within 
$\sim$1\arcsec\ of each other in the central core and are the brightest members
of the cluster. They are believed to be massive main sequence stars which, like
the bright stars in R136a, drive very strong winds \citep{dekoter,crowther}.
NGC~3603 has also been shown to be a seat of active star formation outside
the main cluster region (e.g. \cite{brandner2}). In particular, three distinct
dense cometary nebulae were detected there using HST and VLT: Are these ultra
compact \HII\  regions (UCHR) or ProPlyDs ?

UCHRs are small ($<$0.1~pc) nebulae, {\em internally} photoionized by a deeply
embedded, massive star. Even though the embedded stars are very luminous, they
are invisible at optical wavelengths because of the surrounding dust; instead
they show strong far-infrared emission. Their radio continuum emission is 
often associated with OH and H$_2$O masers. The morphology of UCHRs in high 
resolution radio continuum images ranges from spherical, cometary, core-halo,
shell, to irregular \citep{wood}. 

In contrast, a proto-planetary disk (ProPlyD) is a phenomenon describing a 
low-mass, young stellar object (YSO) with a circumstellar disk, embedded in
a dense (neutral and ionized) envelope that is being {\em externally} 
photo-evaporated by the ultraviolet radiation from one or more massive stars. 
These low-mass stars are not able to ionize their surroundings significantly. 
The disks in these systems are either observed directly (as in Orion where the
disks are seen directly or in silhouette against the bright background nebula
\citep{bally}), or are inferred because in almost all of the ProPlyDs (young)
low-mass stars are visible at optical or near-infrared wavelengths (see e.g.
\cite{stecklum}). At low angular resolution ProPlyDs look like UCHRs, but at
high resolution, optical and near-infrared images reveal their different
nature. So most ProPlyDs would have been classified as UCHRs prior to those
observations. In some cases the distinction between the two classes remains 
difficult. Given the fact that there are about four times more UCHRs than
expected \citep{churchwell} from star formation rates and the potential
problems identifying host stars for UCHRs (e.g. some may not have enough 
FIR flux to account for an internal OB star), it may well be that a large
fraction of catalogued UCHRs are mis-identified ProPlyDs. Interferometric radio
continuum observations of ProPlyDs are needed to determine their structure, 
spectral indices and thus the nature of their emission, as well as mass loss 
rates and extinctions when compared to measured \Ha\ fluxes.

ProPlyDs were first identified in the Orion
Nebula\footnote{ There are a variety of different terms used in the
literature for  describing the knots of ionised gas in M\,42 (Orion): e.g. CKs
(cometary knots), PIGs (partially ionized globules), EIDERS (external ionised 
(accretion) disks in the environs of radiation sources), and ProPlyDs 
(proto-planetary disks); for a summary see \cite{mccullough}. Another 
expression being used is EGGs (for evaporating gaseous globules) describing
the objects found in M\,16 (Eagle Nebula). The cometary knots found in the 
Helix Nebula are compact globules and very different from ProPlyDs as they
contain no stars.}, where over 150 of them are known \citep{odell}. Two
ProPlyDs have been identified in more distant nebulae: one in NGC~2024 by 
\cite{stapelfeldt} and one, G5.97--1.17, in the Lagoon Nebula by
\cite{stecklum}. The three {\plike} objects, hereafter referred to P1, P2
and P3, in NGC~3603 \citep{brandner2} are the biggest, youngest and most 
massive ones found {\it{so far}}. They are also the most distant known.
These emission nebulae are clearly resolved in the HST/WFPC2 observations, and
share the overall morphology of the ProPlyDs in Orion. All three nebulae are
rim-brightened and tear-drop shaped with the tails pointing away from the
central ionizing cluster. 

According to \cite{brandner2} the brightest object (P1), which has a 
projected distance of 1.3~pc  from the cluster, has the spectral (excitation) 
characteristics of a UCHR. Optical spectra reveal the presence of an 
underlying, heavily reddened continuum source, which is also confirmed by
near-infrared VLT/ISAAC observations.  The WFPC2 observations show that only
the outermost layer is ionized whereas the interior is neutral. The morphology
of P1 is described as a heart-shaped head with a collimated structure in
between, which can be understood as the superposition of two individual
ProPlyDs. In contrast, P2 and P3, located at projected distances of 2.2~pc
and 2.0~pc from the stellar cluster, respectively, show approximately
axisymmetric morphologies. No embedded disk or central star has been detected
so far in any of these nebulae, preventing a clear identification as 
proto-planetary disks. The optical point source (see Fig.~2c) close to P3 is
probably not physically linked to the nebula \citep{brandner2}. The {\plike}
structures in NGC~3603 are about two orders of magnitude fainter than typical
UCHRs, but have a typical extent of 9000 AU with tails extending to 21000 AU,
much larger in size than the ProPlyDs in Orion.

Recent 3.4~cm radio continuum and recombination line measurements of NGC~3603
by \cite{depree}, which were focussed on abundance measurements and the bright
continuum emission from the ionized gas in this region, have an angular 
resolution of $\sim$7\arcsec\ and a $5\sigma$ sensitivity of 55 mJy. By 
obtaining high sensitivity and high angular resolution 
($\sim$1\arcsec--2\arcsec) {\ATCA} observations we primarily aimed to study
the radio emission of the winds from many of the early-type stars in
the cluster, as well as to detect and resolve the {\proplyd}s and other
gaseous regions in the cluster periphery. This paper will focus on the {\plike}
sources\footnote{Throughout this paper we use the term ``ProPlyD-like'' for 
these cometary-shaped objects in NGC~3603, since the identification of these
objects as true ProPlyDs is premature, owing to the lack of a clear detection
of a central disk or star.} only, which have been detected and are shown to be
clearly resolved with the \ATCA. A subsequent paper containing a detailed 
study of the whole NGC~3603 region based on our \ATCA\ observations will 
follow.

\section{Observations and Data Reduction}
Radio continuum observations of NGC~3603 were made with the Australia Telecope
Compact Array (\ATCA) in the 6A, B, C and D configurations in five observing runs
in February, April, June, September and November  2000 with a total of 60 hours assigned observing
time. The observing frequencies were 4.8~GHz (6~cm) and 8.64~GHz (3~cm) with a
bandwidth in each case of 128 MHz. Detailed observing parameters are given in
Table~1. The three data sets were combined, reduced and analysed in MIRIAD 
using standard procedures. The full Stokes parameters were measured. The flux
density scale was calibrated using observations of the primary calibrator, 
1934--638, assuming flux densities of 2.84 and 5.83 Jy at 3 and 6~cm,
respectively.

A primary beam correction was carried out. To obtain high angular resolution
and filter the extended emission to emphasize the small scale structure we
used uniform weighting of the {\it{uv}} data while omitting the shortest
baselines, up to $25\,{\rm k}\lambda$, respectively.
We find that for a $25\,{\rm k}\lambda$ cutoff all bright extended nebula emission
is removed from the {\proplyd}s even at 6~cm. Fig.~1 shows an example of a combined 3~cm radio map where
a $uv$ cut of $10\,{\rm k}\lambda$ was chosen. A zoom on the three {\proplyd}s is
shown in Fig.~2. For the 6~cm maps a $uv$ cut of $25\,{\rm k}\lambda$ was used. The
average r.m.s. in these combined radio maps was approximately 0.2 mJy at 3~cm
and 0.4 mJy at 6~cm, but varies throughout the map. The r.m.s. levels near
the {\proplyd}s, which lie in rather empty regions, were $\sim 0.1$ mJy
at 3~cm and $\sim 0.2$ mJy at 6~cm.
The dynamic range in the 3 and 6~cm images is about 1:100. Fitting a
2D-Gaussian to the dirty beam gives FWHM values of 0\farcs94 $\times$ 0\farcs77
to 1\farcs06 $\times$ 0\farcs87 at 3~cm and 1\farcs53 $\times$ 1\farcs23
to 1\farcs83 $\times$ 1\farcs49 at 6~cm depending on the $uv$ cut employed.
In Fig.~2 we have restored the
cleaned maps with a circular Gaussian beam of width 1\arcsec\ at 3~cm and
2\arcsec\ at 6~cm.

To create spectral index maps between 3 and 6~cm we convolved the 3~cm map 
with a Gaussian appropriate to achieve the resolution of the beam
FWHM at 6~cm. The spectral index maps were then derived from the 6~cm and
convolved 3~cm maps using a clip value of 1.0~mJy beam$^{-1}$, corresponding
to approximately $10\sigma$ at 3~cm and $5\sigma$ at 6~cm (statistical errors). Error maps for
the spectral index maps were calculated using error propagation.--- We note,
however, that spectral indices of extended sources as obtained from interferometric
data have to be regarded with caution.

\placetable{table1}

\section{Results}
The three {\plike} sources found with the
\ATCA\ at 3 and 6~cm are clearly  resolved, showing a head-tail extent of
$\sim$4\arcsec\ (see Fig.~2). P3  shows the most pronounced head-tail
structure with a 3~cm flux density ratio  between head and tail of about 10:1.
The tail is very well defined and at least 2\arcsec\ long, pointing away from
the central star cluster. Unfortunately, P3 is rather faint in the
low-sensitivity HST broad band  image shown by \cite{brandner2}; it is located
outside the region of their high-sensitivity \Ha\ image.

\subsection{ProPlyD structure and fluxes}

The three \proplyd s have a cometary or head-tail shape\footnote{The silhouette
of ProPlyDs has variously been referred to as cometary, tadpole, tear-drop or 
simply head-tail shaped.} at cm-radio wavelengths, matching their appearance
in optical images (see Fig.~2).

Integrated radio flux densities and peak fluxes are derived by summing over the whole
{\proplyd}s, and by fitting circular Gaussians to each object.
Statistical flux uncertainties are estimated as map r.m.s. times number of beamsizes
covered by the source. Dependent on the method of flux determination, beam size  and $uv$ cut
employed to the maps (5, 10, 15, 20, 25 k$\lambda$) the resulting flux densities
vary. The standard deviation of this variation is considered as systematic uncertainty.
We found that the systematic uncertainties are always larger than the statistical
error. In Table~2 we report the average fluxes and their uncertainties
of all 3 nebulae, together with their average spectral indices derived from the integrated
fluxes.

Circular Gaussians\footnote{We find that elliptical
Gaussians instead of circular ones would fit too much tail flux to the head.}
are used to at least attempt disentangling the head and tail structures in these cometary-shaped
objects.
The radio heads of the {\proplyd}s are typically 2\arcsec--3\arcsec\ in 
diameter while the total head-tail extent is $\sim$4\arcsec. P3 shows the most
pronounced head-tail structure of the three (most likely
due to the viewing angle) and the smallest head (diameter $\sim$2\arcsec)
among all three {\proplyd}s, with the largest peak flux at 3~cm while the
integrated head flux is comparable. Its head-to-tail flux ratio is about 
10:1, whereas in the other two {\proplyd}s part of the tail may be confused
with the head flux density.

Both, P1 and P2, appear inclined with respect to the plane of the sky, with P1
inclined at a larger angle than P2. This can be seen in Fig.~2 where we have 
indicated the projected direction to the ionizing source. Fig.~3 shows the 
residual radio continuum image of P1 after removing a single
circular Gaussian component.  The residual map of P1 shows two excess
emission sites, one corresponding to the tail, and a second one north of it.
Recent \Ha\ images indicate that P1 is composed of two separate cometary-shaped
objects \citep{brandner2}. We interpret the northern excess as the second
head. A two component Gaussian fit to P1 at 3~cm gives peak fluxes of 4.6 mJy
beam$^{-1}$ and 1.4 mJy beam$^{-1}$, and integrated flux densities of
9.9 mJy and 5.7 mJy for the two heads, with a
separation of the two peaks of 1\farcs3. A two component model could not be
applied to the lower resolution 6-cm data.

The tails of all three {\proplyd}s are directed away from the star cluster,
approximately coincident with the tails in \Ha\ within the positional
uncertainty of the radio/optical pointings (see Fig.~2; note the
shift of 0\farcs5 in declination and right ascension of the HST-image which we have employed in
these figures). The uncertainty of the radio position is $\sim$0\farcs15 while
HST has a pointing uncertainty of $<1$\arcsec. 

The present data were obtained in all four Stokes parameters. No polarization
signal above 3$\sigma$ was detected.

Possible flux variability on timescales $\ga$ days expected from e.g. flaring
pre-main sequence (PMS) stars with non-thermal radio spectra, cannot be tested
with the present data.

\placetable{table2}

\subsection{Spectral indices of the ProPlyDs}
The radio continuum fluxes (see Table~2) are a factor of 9 to 21 larger
than those predicted by \cite{brandner2} assuming optically-thin thermal 
bremsstrahlung as the radio emission mechanism \citep{mccullough}, which
scales with the (extinction corrected) \Ha\ flux. The neutral part of the
envelope of the {\proplyd} may attenuate \Ha\ photons from the far side
of the object depending on the viewing angle. However, \cite{mccullough}
have estimated this effect to be less than 25\% of the total \Ha\ light for
an isotropic ensemble of ProPlyDs. Furthermore, the reasonable coincidence
of the sizes of the {\proplyd}s in the radio and optical band suggests that
this effect is not important here. For P1 and P3 the integrated
fluxes at  6~cm are higher than that at 3~cm, whereas similar
flux densities were predicted.
The average spectral index of each object, shown in Table~2, is
calculated on the basis of the integral fluxes at 3~cm and 6~cm in
this table. Modeling the uv distribution of the 6~cm map to match the uv distribution
at 3~cm gives similar results.

The average spectral indices of P1 and P3 are
negative (P3 only marginally), between $\alpha = -0.5\pm0.2$ and $-0.3\pm0.2$
($S_\nu \propto \nu^\alpha$), indicating non-thermal emission, while the
spectrum of P2 ($\alpha \approx -0.1\pm0.2$) is in agreement
with the expectation from optically thin thermal bremsstrahlung. 

Hydrodynamic simulations predict spectral indices of $\alpha \approx 0$
near the ionization fronts, which can increase up to $\alpha \approx 0.6$ when
considering in addition a hydrodynamically collimated jet (S.~Richling, private
communication). Negative spectral indices cannot be explained by this model.

Because of the high resolution and sensitivity in the radio maps, we were able
to generate spectral index maps for all three {\proplyd}s.
Fig.~4 shows examples of the spectral index maps. Here we used the radio maps
shown in Fig.~2 and constrain ourselves to
brightness densities greater than 1.0~mJy\,beam$^{-1}$, which corresponds to
roughly 5~$\sigma$ at 6~cm and 10~$\sigma$ at 3~cm (based on statistical error estimates).
The radio emission zone within each {\proplyd} appears to
be inhomogeneous. The spectra in at least two of the sources appear steeper in the tail than in the
head, with a tendency of positive spectral indices located towards the region
facing the ionizing star cluster.

\subsection{Extinction towards the ProPlyDs}

\cite{melnick} have derived the reddening towards NGC~3603 from UBV photometry,
and found $A_V \approx 4-5$~mag for the cluster core. However, as can be seen
in their data, the visual extinction is not uniform across the whole star 
forming region, but increases significantly with distance from the star 
cluster, mainly towards the North and the South. This is confirmed by recent
VLT/ISAAC data on the basis of 4750 stars, showing that the extinction can be
as high as A$_{\rm{V}} \approx 14$~mag (Brandner, private communication). For
example, star MTT~68, close to P3, has A$_{\rm{V}} \approx 6.2$~mag, whereas
star MTT~81 near P1 has A$_{\rm{V}} \approx 4.7$~mag, and star MTT~79 close to
P2 has extinction A$_{\rm{V}} \approx 5.2$~mag. 
Here we use $R = A_{\rm{V}}/E_{\rm{B-V}} \approx 3.2$ as used by Melnick et
al. (1989) for NGC~3603. Thus, it appears likely that P2 and P3 suffer from
stronger foreground extinction than the stars in the cluster center which
have on average A$_{\rm{V}} \lesssim 4.5$~mag. 

Using the ratio between radio flux density and \Ha\ flux as given by 
McCullough et al., (1995; their equation 5) and assuming that the radio flux density
at 3~cm is dominated by thermal bremsstrahlung, we predict the \Ha\ flux using
our 3~cm radio continuum data. The predicted fluxes are then compared to the
\Ha\ fluxes measured by \cite{brandner2} and used to derive lower limits for
the extinction, $A_{\rm H\alpha}$. (If (gyro-)synchrotron radiation dominates 
the radio emission, which comes from an emitting region as specified in 
Sect.~4.4, the derived extinction $A_{\rm H\alpha}$ would be higher, with the
exact value depending on the magnetic field.)
Table~3 shows the calculated fluxes (assuming $T_{\rm e} = 10^4$~K) and
extinctions. Note the uncertainty in comparing different areas for which the
\Ha\ fluxes and the radio continuum fluxes have been obtained. \cite{brandner2}
give the \Ha\ surface brightness of the {\proplyd} heads, measured for a
0\farcs5 aperture radius, whereas we quote peak fluxes per beam where the 
beam size is $\sim$1\farcs0$\times$1\farcs0. For $A_{\rm H\alpha} \approx 0.85
A_{\rm{V}}$, we find A$_{\rm{V}} \approx 6.0$ for P1 and A$_{\rm{V}} \approx
6.7$ for P2. We find thus significant excess extinction for both {\proplyd}s, which
may be intrinsic to the {\proplyd}s themselves, otherwise the extinction 
estimate derived from the stars is underestimated.

\placetable{table3}

\subsection{ProPlyD densities and mass-loss rates}

Estimates for the electron/ion density inside the heads can be
obtained from our radio data if thermal bremsstrahlung dominates the radio
emission. This is likely to be the case for P2 and maybe P3, while for P1
the derived values should be considered as upper limits. The
bremsstrahlung intensity in the radio domain for a thermal, ionized H-gas of
constant density in a sphere is \citep{mezger}
$$
S_{\rm{radio}} \approx 1.9 D_{\rm{kpc}} f_{\rm{th}} \theta_{\rm{sec}}^3
N_{e,4} N_{\rm{ion},4} \nu_{\rm{GHz}}^{-0.1} T_4^{-0.35} \rm{mJy},
$$
where $D_{\rm{kpc}}$ is the source distance in kpc, $\theta_{\rm{sec}}$ is the 
apparent radius of the source in arcsec, $N_{e,4}$ and $N_{\rm{ion},4}$ are the
densities of the thermal electrons and ions, respectively, in
$10^4$~cm$^{-3}$, $\nu_{\rm{GHz}}$  is the observing frequency in GHz and
$T_4$ is the electron temperature of the ionized gas in $10^4$~K.
$f_{\rm{th}}$ is the ratio of thermally emitting to total volume of the
source, and is introduced to account for the possibility of smaller emitting
volumes than the apparent extent of the source. This is likely because only
the outermost layers in ProPlyDs are believed to be ionized. We estimate
the ProPlyD heads $\sim$ 2-3\arcsec\ in diameter. For $T_4 = 1$ and $N_{e} \approx
N_{\rm{ion}}$ we estimate the electron density by comparing the expected radio
flux with the observed 3~cm ProPlyD head flux densities, and find $N_{e} =$(9,
7 and 13)$\cdot 10^3 f_{\rm{th}}^{-1/2}$~cm$^{-3}$ for P1, P2 and P3,
respectively. (The corresponding emission measures are EM = 5.9, 3.5 and 9.3
$\times 10^6 f_{\rm{th}}^{-2/3}$~cm$^{-6}$~pc.)  If the emitting volume is
smaller than the total source volume, the electron density and emission
measure increase accordingly. From spectral considerations the
emission must be optically thin, i.e. free-free absorption may set in
at frequencies lower than 4.8~GHz. Thus
values of EM cannot exceed $\sim$ several $10^7$~cm$^{-6}$~pc, in
order to avoid  free-free absorption at frequencies above 4.8~GHz.
This constrains the size of the thermally emitting volume of the 3 {\proplyd}s
to at least 4\%, 2\% and 9\% of the total source volumes, and consequently puts
limits on the thermal electron densities in the sources: 4.5, 4.9 and 4.3
$\cdot 10^4$~cm$^{-3}$ for P1, P2 and P3, respectively. Note that these
numbers should be considered as upper limits if the observed radio flux is
dominantly of non-thermal origin. These numbers are in
reasonable agreement with $N_e \geq 10^4$~cm$^{-3}$ derived from [SII] line
ratios by  \cite{brandner2}. For the Orion ProPlyDs densities of order 
$10^5-10^6$~cm$^{-3}$ have been found \citep{henney}. Note that our derived
numbers suffer from uncertainties due to the unknown geometrical projection
angle of the {\proplyd}s, and hence the true source diameters. 

Mass loss from the {\proplyd}s occurs through external heating of the envelope by 
the far ultraviolet (FUV) radiation field from hot stars, especially from the
core of NGC~3603, and subsequent evaporation. Knowing the electron density, the
mass-loss rate in a wind is given by $\dot M = 4\pi R^2 N_e \mu m_H v_W$, with
$R$ the system radius, $m_H$ the ion mass and $v_W$ the evaporation velocity.
Assuming a pure hydrogen gas, $\mu=1.3$, we find $\dot M/v_{20} \approx
(5-10) \times 10^{-5}$\Msun year$^{-1}$ for the heads of all three
{\proplyd}s, where $v_{20}$ is the wind velocity in units of 20 km s$^{-1}$.
\cite{brandner2}  found evaporation flow velocities of order 10--25 km
s$^{-1}$. The derived upper limits for the mass-loss rates are of the
same order as estimated by \cite{brandner2}, and about two orders of magnitude
higher than the typical  values for the Orion ProPlyDs.

A rough estimate of the mass reservoir of the NGC~3603 {\proplyd}s can then 
be derived by multiplying the mass loss rate with an estimated evaporation 
time scale of $\sim 10^5$ years \citep{brandner2}. This gives a mass for the
{\plike} objects of order 1--10\Msun, 10--1000 times larger than for the Orion
ProPlyDs.  The radius of the putative disks in these massive objects can be
estimated using the evaporation model of \cite{johnstone} which considers the
mass reservoir in a disk-like shape. We find an approximate disk radius of 
2000~AU ($\approx 0.3$\arcsec) for all three NGC~3603 {\proplyd}s:
\cite{brandner2} found $\sim$3400~AU from hydrodynamical simulations.

\section{Discussion}
\subsection{Comparison with the ProPlyDs in the Orion and Lagoon Nebulae}

The {\proplyd}s  in NGC~3603 are 20--30 times larger and much more spectacular
than those in Orion, which were detected at 2 and 20~cm with the VLA (see
\cite{mccullough} and references therein; \cite{henney}). 

Orion contains numerous dense clumps, originally classified as UCHRs, but now
known to be ProPlyDs. Most Orion ProPlyDs have central disks, sometimes seen
only in silhouette against the background light. On recent HST-images dozens of
jets powered by young stars embedded in ProPlyDs have been found \citep{bally}.
All Orion ProPlyDs detected at radio wavelengths are non-variable thermal 
emitters with spectral indices between --0.1 and 0.2 \citep{felli}, and located
within 0.04~pc of the dominant ionizing O-star ($\theta^1$ Ori C). The 
non-thermal Orion radio sources show strong variability, and are mostly
associated with visible pre-main sequence (PMS) stars. Their non-thermal 
spectra are usually explained as due to stellar flaring activity. 

G5.97--1.17 ($D$ = 1.8~kpc) was originally classified as an ultracompact \HII\
region, but is now thought to be a ProPlyD at $\sim 5000$ AU projected distance
from the O7 star Herschel 36 in the center of M8, the Lagoon Nebula
\citep{stecklum}. ESO near-infrared, HST optical and VLA
radio continuum observations indicate that G5.97--1.17 is a young star
surrounded by a circumstellar disk that is likely being photo-evaporated by Her
36, similar to the ProPlyDs in Orion.  The previous hypothesis
suggested that  G5.97--1.17 is a UCHR intrinsically ionized by
an embedded B0 star. However, \cite{stecklum} showed that
G5.97--1.17 is predominantly externally ionized, and thus the spectral type of
the embedded star should be later than B5.  The \Ha\
flux over 0\farcs6 is consistent with the $\lambda=2$~cm appearance. Optical
and near infrared (NIR) continuum images show the central star is displaced from
the peak of the bow shock by 0\farcs125. NIR photometry of G5.97--1.17 revealed
that its central star is extremely red, which cannot be explained by
extinction laws that use spherical matter distributions. This supports the
idea of the circumstellar matter surrounding the central star of G5.97--1.17
being arranged as a disk.
The cm-radio spectrum of G5.97--1.17 appears flat,
which is interpreted as optically thin free-free emission
\citep{wood,doherty}. No OH or H$_2$O masers, which are often associated
with UCHRs, are known in G5.97--1.17. 

In comparison, the NGC~3603 {\proplyd}s appear rather peculiar
and spectacular. Not only are they the largest and most massive ones found so
far. Exposed to an extremely strong FUV radiation field, they also suffer the
largest mass losses, though their distances from the ionizing source 
are larger than for the Orion or Lagoon Nebula ProPlyDs. No embedded
source nor disk has been found for any of the NGC~3603 {\proplyd}s. 
At cm-wavelengths they are so far the only {\proplyd}s showing non-thermal
radio emission.

The spatial distribution of the NGC~3603 {\proplyd}s is also peculiar. They
are not only distributed apparently along a straight line (see 
Sect.~4.3 for further discussion), but all three also lie clearly to the north
of the ionized gas region and avoid the region north-east and south-west
from the stellar cluster.
This is in contrast to the Orion ProPlyDs which
do not appear to be located in preferred regions of the nebula (e.g.
\cite{bally}). Since sequential star formation proceeding from north to south
seems evident in NGC~3603 (e.g.  \cite{depree}), this suggests that the
{\plike} structures in this region might have emerged just recently.

\placetable{table4}

\subsection{External versus internal ionization}

A possible way to prove that the {\proplyd}s have indeed been ionized {\it
externally} by the star cluster is to show that the observed radio fluxes 
for the {\proplyd}s are consistent with the number of ionizing photons they 
receive from the cluster stars. For simplicity we assume thermal radio 
emission, and note that this leads to an upper limit for the number of
ionizing photons needed to produce the observed radio flux density. In this
case the brightness temperature is directly proportional to the number of
ionizing photons. Because NGC~3603 contains a large number of stars that 
are extremely hot and luminous they provide an ionizing flux that is several
orders of magnitude larger than that in Orion or M~8 (see Table~4). The 
expected brightness temperatures at the projected distances of 1.3~pc, 
2.2~pc and 2.0~pc of the {\proplyd}s from the ionizing source due to a Lyman 
flux of $10^{51}$photons s$^{-1}$ \citep{brandner2} are 200~K, 70~K and 90~K
at 3~cm assuming $T_4=1$. This is only slightly higher than our estimates 
derived from the {\ATCA} 3~cm maps of 40~K, 30~K and 70~K for the three
{\proplyd}s, respectively. We can also compare the number of ionizing photons
required to deliver the observed flux densities at the location of the
{\proplyd}s, and compare this number with the number of Lyman photons they
receive from the star cluster. Noting that the {\proplyd}s subtend a fraction
of (6, 3 and 2)$\cdot 10^{-5}$, of the total solid angle seen by the cluster
stars, they thus receive (6, 3 and 2)$\cdot 10^{46}$~photons~s$^{-1}$, 
respectively.  On the other hand, the required number of ionizing photons 
at the location of each {\proplyd} to deliver the observed radio flux density
is about $2\cdot 10^{46}$~photons~s$^{-1}$ in all three cases. These
comparisons show that the number of required Lyman
photons is slightly lower than provided by the star cluster, and thus would
even allow an additional dust attenuation and/or, more likely, longer linear
distances between the {\proplyd}s and the ionizing source in the external
ionizing scenario. Significant ionizing power from within the {\proplyd}s
appears therefore unlikely, which puts the spectral type of any central stars
to later than $\sim$B1, and hence mass to below $\sim 10$\Msun.

\subsection{Do the {\plike} objects host disks ?}

The present radio data as well as the recently published HST/VLT-images do not
provide any direct evidence that the observed {\plike} objects in NGC~3603 host
any (proto-)stellar objects or disks. To a limiting $K_s$-magnitude of 18, no
circumstellar disk or central stellar object in any of the three {\proplyd}s
has been detected by HST+VLT observations. Instead the cometary-shaped nebulae
may rather be massive emission regions photo-ionized by the radiation from the
star cluster. \cite{mellema} have presented model calculations for such clumps,
called FLIERs or ANSEA which are commonly located along the major axis towards
the ionizing source. Indeed, all three {\proplyd}s apparently lie approximately
along one straight line, with the outflow source, Sher~25, located roughly 
midway between P1 and P3. Although the bipolar outflow from Sher~25 appears 
not to point in the direction of any of the three {\plike} sources, the
ring-like structure around Sher~25, roughly perpendicular to its jet, has been
shown to expand with a velocity of $\sim 30$ km s$^{-1}$ \citep{brandner1},
and could in principle be related to the origin of the {\plike} clumps. 
In addition, models such as the FLIER model which are not based on disk-like
mass distributions typically give short evaporation times, inconsistent with
the distance of the {\proplyd}s from Sher~25 (W.~Brandner, private
communication) and the approximate date of
the mass-loss event about 6630 years ago in Sher~25 \citep{brandner1}. Thus,
the existence of such features at large distances from the cluster centre
appears to favor models where most of the material is concentrated in
compact structures such as disks. The actual presence of disk-like structures
is however not confirmed.

\subsection{The possible origin of the non-thermal emission}

The significantly non-thermal spectrum ($\alpha \ll 0, S_\nu \propto \nu^{\alpha}$)
from at least one of the three NGC~3603 {\proplyd}s (P1; P3 seems to be marginally
non-thermal) appears to be steepest ($\alpha < -0.5$) in the tail,
slightly flatter ($\alpha = -0.5$ to 0.1) in the head, and have a
tendency of positive spectral indices ($\alpha > 0$) located towards the region
facing the ionizing star cluster rather than away from it. This is not expected from 
non-thermal radio emission from a wind--wind collision region. The non-thermal
radiation must originate rather from inside the {\proplyd}s, either from their
putative central stars hidden behind thick material, the surrounding disks or
the envelopes. 
A stellar origin analogous to active star coronal emission
seems to be unlikely because of the extended nature of the observed source.
Non-thermal radio spectra could be produced by a population of
energetic particles emitting synchrotron, gyrosynchrotron or gyroresonance
radiation in a magnetic field. 
For gyrosynchrotron radiation
the emission is concentrated at harmonic numbers $s=\nu/\nu_B = 10 - 100$
(where $\nu_B \approx 2.8\cdot 10^{-3}
B_G$~GHz is the gyrofrequency, $B_G$ is the magnetic field strength in Gauss),
and thus magnetic fields strengths of order $B \approx 20-200$~G are required.
These high field values cannot be present in extended sources, which leaves
synchrotron radiation as the only plausible emission mechanism.

Disk or envelope, whatever the ultimate source might be, the negative spectral
indices suggest optically thin non-thermal processes as the dominant radiation
mechanism in P1, and maybe P3. One could argue that the shock  created by the
evaporation flow predicted in ProPlyD models could accelerate charged
particles within the objects. Other possibilities may be inflow-outflow
activity causing shocks during the early stage of star formation, compression
and reconnection of magnetic fields in the collapsing envelope, or magnetic
reconnection in star--disk interactions in Young Stellar Objects leading to
production of energetic particles. Magnetic fields are known to exist in
Molecular Clouds \citep{crutcher}, and therefore it appears plausible that gas
clumps still contain ``fossil'' fields. In a magnetized plasma, acceleration
due to particle collisions and subsequent bremsstrahlung can often be
negligible in comparison with acceleration due to gyration around the field
lines. In place of free-free emission, there is then synchrotron emission from
energetic particles.  The negative radio spectral indices in P1 and P3 are
conveniently explained by non-thermal particle spectra. Because thermal
bremsstrahlung is proportional to $N_e^2 T_e^{-1/2}$ and synchrotron
emission is proportional to $N_e B^\beta$ ($\beta > 0$), the
former dominates if the density is high enough, or if the temperature or field
strength is low enough. Indeed the electron densities found in the NGC~3603
{\proplyd}s are $\sim$two orders of magnitude smaller than in the Orion
ProPlyDs (see Table~4). 
The different energy dependencies of the
two kinds of emission may lead to one dominating at low frequencies and the
other at high frequencies.

A lower limit for a possible magnetic field is given by the Razin-Tsytovich
effect. In order to avoid a Razin-suppression of the radio spectrum above
$\lambda=6$~cm the magnetic field cannot be smaller than $\sim 10^{-5}$~G
for an electron density of $10^4$~cm$^{-3}$ in the emission region. On the
other hand, to avoid gyroresonance absorption above $\lambda=6$~cm the field
strength
cannot be higher than $10^3$~G. The equipartition magnetic field, the minimum
possible value in flare loops, for a $T_e = 10^4$~K particle distribution with
density $10^4$~cm$^{-3}$, is $\sim 10^{-3}$~G. 
Note that synchrotron photons of energy 4.8 GHz in
a $10^{-5}$ G field must have been produced by $\sim 10$ GeV electrons.
Efficient particle acceleration may occur, e.g., at the ProPlyD surface,
from flow amplification of even very weak ambient magnetic fields followed by local
reconnection events (e.g. \cite{larosa}).

The flux density from a homogeneous sphere with diameter $l$ due to synchrotron
radiation in a uniform field and a low energy cutoff $E_{\rm{min,MeV}}$ in
units MeV of a $E^{-\alpha_e}$ particle distribution with
$\alpha_e=2$ spectral index\footnote{Shock acceleration typically predicts a
$E^{-2}$ power law, which we use as a reasonable assumption here.}  
is given by (e.g. Dulk 1985)    
$$
S_{\nu,\rm{syn}} \approx 5 \times 10^{18} N_{e,nt,4}
B_{G}^{1.5} E_{\rm{min,MeV}} l_{\rm{pc}}^3 D^{-2}_{\rm{kpc}} \nu_{GHz}^{-0.5}
\,\,\, {\rm{mJy}}\,. 
$$
The non-thermal electrons might have been energized by e.g. shock
acceleration or magnetic reconnection events, out of the pool of thermal
particles, or they may be Galactic cosmic rays penetrating the
starforming region. In the latter case $N_{e,nt}\approx 10^{-2} \ldots
10^{-1}$cm$^{-3}$ and $\alpha_e\approx 2.4$, and thus $B \approx 10^{-3}$~G
(for $E_{\rm{min,MeV}}=1$). For the
former case, because energy is conserved, the production efficiency of the
high energy particles is related to the properties of the
thermal background particles, and vice versa. Realistic non-linear diffusive
particle acceleration scenarios typically yield $N_{\rm{e,nt}}/N_{\rm{e}}\sim
10^{-5}-10^{-1}$ (i.e. the fraction of total
electrons which end up with high energies; e.g. \cite{ellison}), while
magnetic reconnection might be as efficient as $N_{\rm{e,nt}}/N_{\rm{e}}\sim
0.01$ for the present densities and temperature \citep{tandberg}. 
Since the thermal electron density $N_e$ cannot be significantly smaller
than $\approx$ $10^4$cm$^{-3}$ (derived from [SII] line
ratios; see Sect.3.4), and $N_{\rm{e,nt}} \leq N_e$ if thermal and non-thermal
electrons have the same source, we find that the size of the
actual magnetized volume emitting non-thermal radiation must be smaller
than the appearance of the {\proplyd} on the sky for this case. If
particle trapping is at work,  $N_{\rm{e,nt}}\geq N_{\rm{e}}$ is possible, and
consequently the ratio of emitting to total volume of the {\proplyd} becomes
even smaller.
 
The present data, however, do not allow to quantitatively constrain the
actual size of the source responsible for the non-thermal radiation.

We note that the recent discovery of variable X-ray
emission from some ProPlyDs in Orion by \cite{schulz} support
the idea of a non-thermal emission volume
smaller than the appearance of the {\proplyd} on the sky.
The size of the emitting region was estimated
on the basis of variability arguments to order 1--10 AU.

\section{Summary and Conclusions}
The three massive ($\sim$1--10~\Msun) {\plike} nebulae in NGC~3603, which have
recently been discovered by \cite{brandner2}, have been clearly detected and
resolved with the {\it ATCA} at 3 and 6~cm, with one of them likely to be 
composed of two cometary-shaped objects. Their flux densities are about 10--20
times higher than expected from the \Ha\ measurements, and a non-thermal
average spectrum can be associated with at least one of the three
{\plike} sources. This is the first time that non-thermal radio emission has
been detected from {\plike} sources. Our spectral index maps show that the
emission region is  rather inhomogeneous, with negative spectral indices in
the tail and part of the head whereas positive spectral indices, indicating
thermal free-free  emission, tend to be detected from a small region facing
the star cluster.

We derive upper limits for the mass-loss rates of
$10^{-5}$~\Msun\,year$^{-1}$ and electron densities of $10^{4}$~cm$^{-3}$.
These are in reasonable agreement with estimates from recent HST-images.

We show that synchrotron emission from a magnetized,
relativistic, non-thermal particle population may explain the non-thermal
spectral regions.
Energetic electrons necessary for synchrotron radiation may be
produced through shocks or by magnetic reconnection out of the pool of
thermal particles, or may be Galactic cosmic rays.

A stellar origin of the observed flux densities appears unlikely
considering that the sources are extended. Thus magnetic fields, which have
been neglected in ProPlyD models so far, are possibly associated with the
ProPlyD envelopes or disks, and appear to be crucial in understanding the
physical processes in these {\plike} objects. 

\cite{melnick} have shown that an inhomogeneous dust distribution in NGC~3603
causes increasing extinction with distance from the star cluster. In 
particular, the extinction at the location of the {\proplyd}s is estimated
to lie between $A_V \approx 5-6$ mag. Extinction estimates derived from the
radio- to {\Ha}-luminosity ratio are 1--2 magnitudes higher.

The ultimate discovery of the so-far undetected disks, which are thought 
to be part of all ProPlyDs, would give important hints about the origin 
of the extremely high and non-thermal radio fluxes and the nature of the
cometary-shaped clumps seen in NGC~3603. This may be accomplished in future
mm observations with an upgraded \ATCA.

\acknowledgments
We are grateful to Wolfgang Brandner for providing the HST image of
NGC~3603 to overlay on our radio images, for interesting discussions 
and carefully reading the manuscript.
We thank Jessica Chapman, Jim Caswell, Simon Johnston, Tylor Bourke and
Sergey Marchenko for reading the manuscript and many constructive comments
which improved our paper significantly. 
AM acknowledges a postdoctoral bursary from the Quebec Government.
AFJM thanks NSERC (Canada) and FCAR (Quebec) for financial support.
IRS acknowledges the receipt of a PPARC Advanced Fellowship.

{}

\begin{figure*} 
\caption{3-cm continuum map with a $uv$ cutoff at 10 k$\lambda$ of NGC~3603 overlaid onto the \Ha\ and broad
band HST-image from \cite{brandner2}. Note that in this radio map the small
scale structure is emphasized due to our chosen {\it{uv}}-cutoff in the data
analysis. For smoothing purposes the map was restored with a 2\arcsec$\times$2\arcsec beam size.
The brightest, extended continuum regions correspond to the heads
of the giant gaseous pillars, showing evidence for the interaction of ionizing
radiation with cold molecular hydrogen clouds.
The radio emission from IRS~9, a deeply enshrouded association of protostars at the foot of the
South-Eastern pillar, will be discussed in a forthcoming paper
(M\"ucke et al., in prep.). The three nebulae indicated
are the {\plike} sources; their tadpole shape is shown in Fig.~2.
The contour levels at 3~cm are -0.8, 0.8, 2.0, 4.0, 6.0, 12.0, 24 and 48
mJy\,beam$^{-1}$. The beam is shown inside the box at the bottom left corner.
North is to the top and East to the left.  \label{fig1}}
\end{figure*}

\clearpage

\begin{figure*} 
\begin{tabular}{ll}
\rotatebox{-90}{\includegraphics[width=4.4cm]{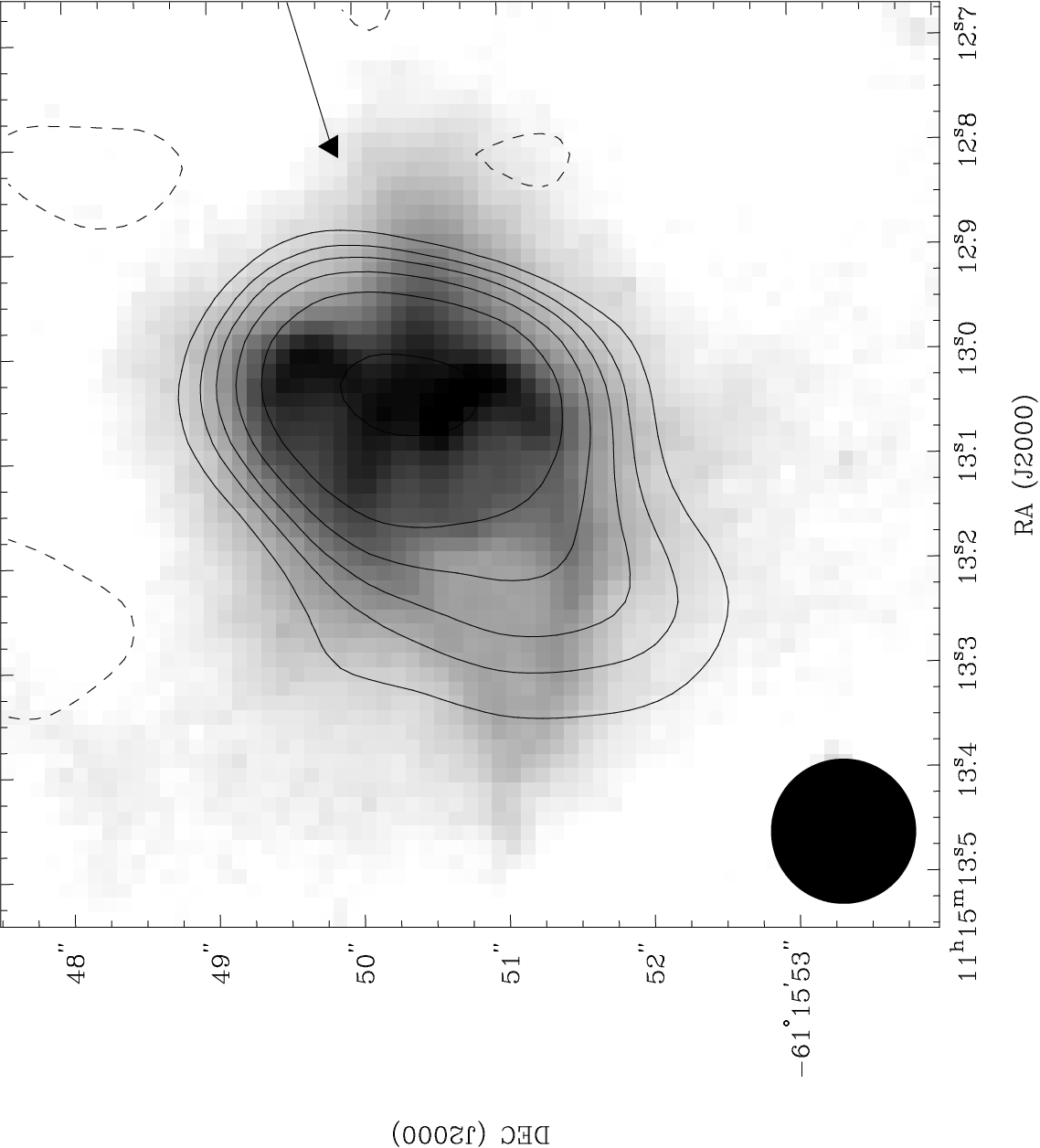}}
\rotatebox{-90}{\includegraphics[width=4.4cm]{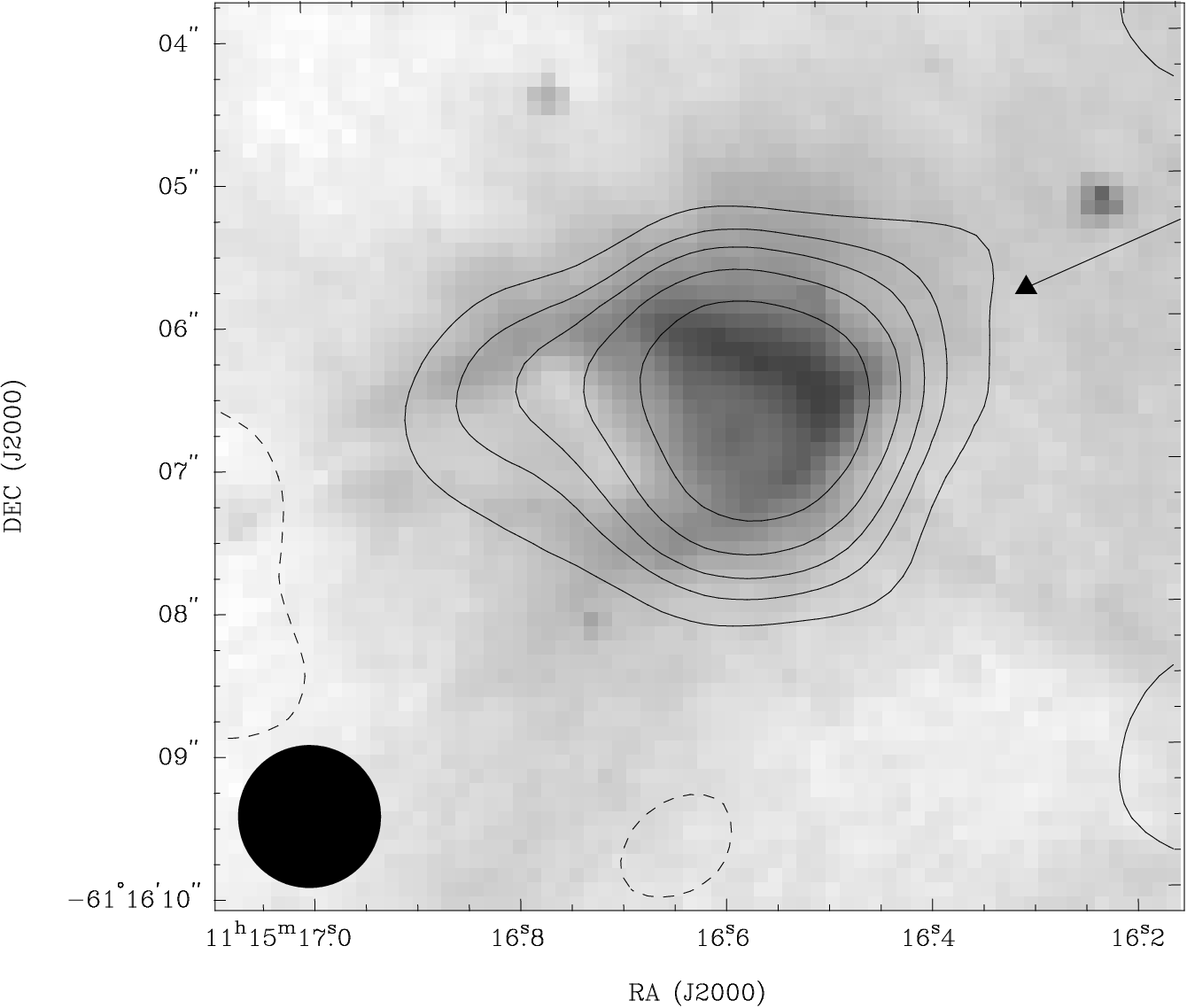}}
\rotatebox{-90}{\includegraphics[width=4.4cm]{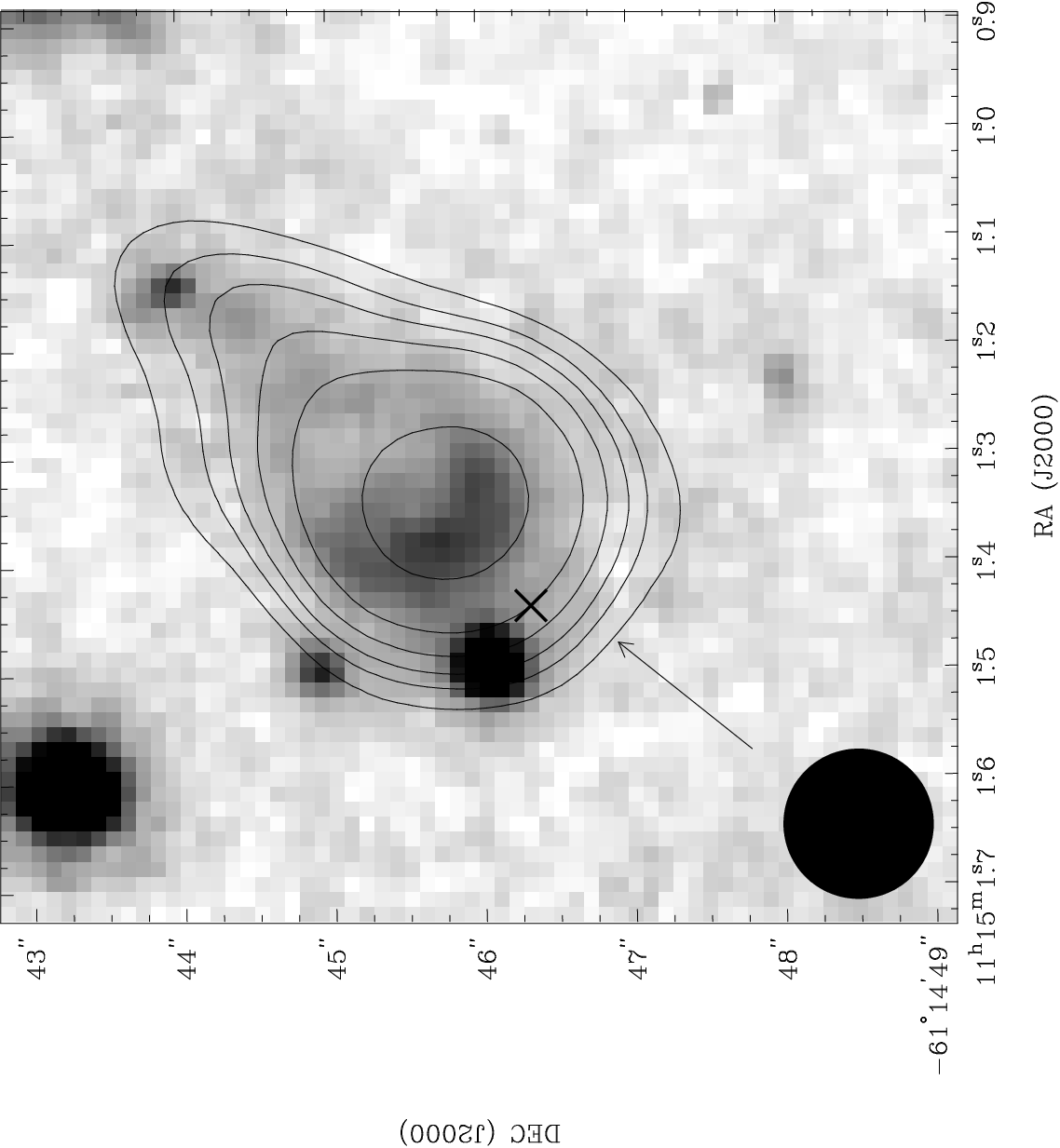}}
\end{tabular}
\begin{tabular}{ll}
\rotatebox{-90}{\includegraphics[width=4.4cm]{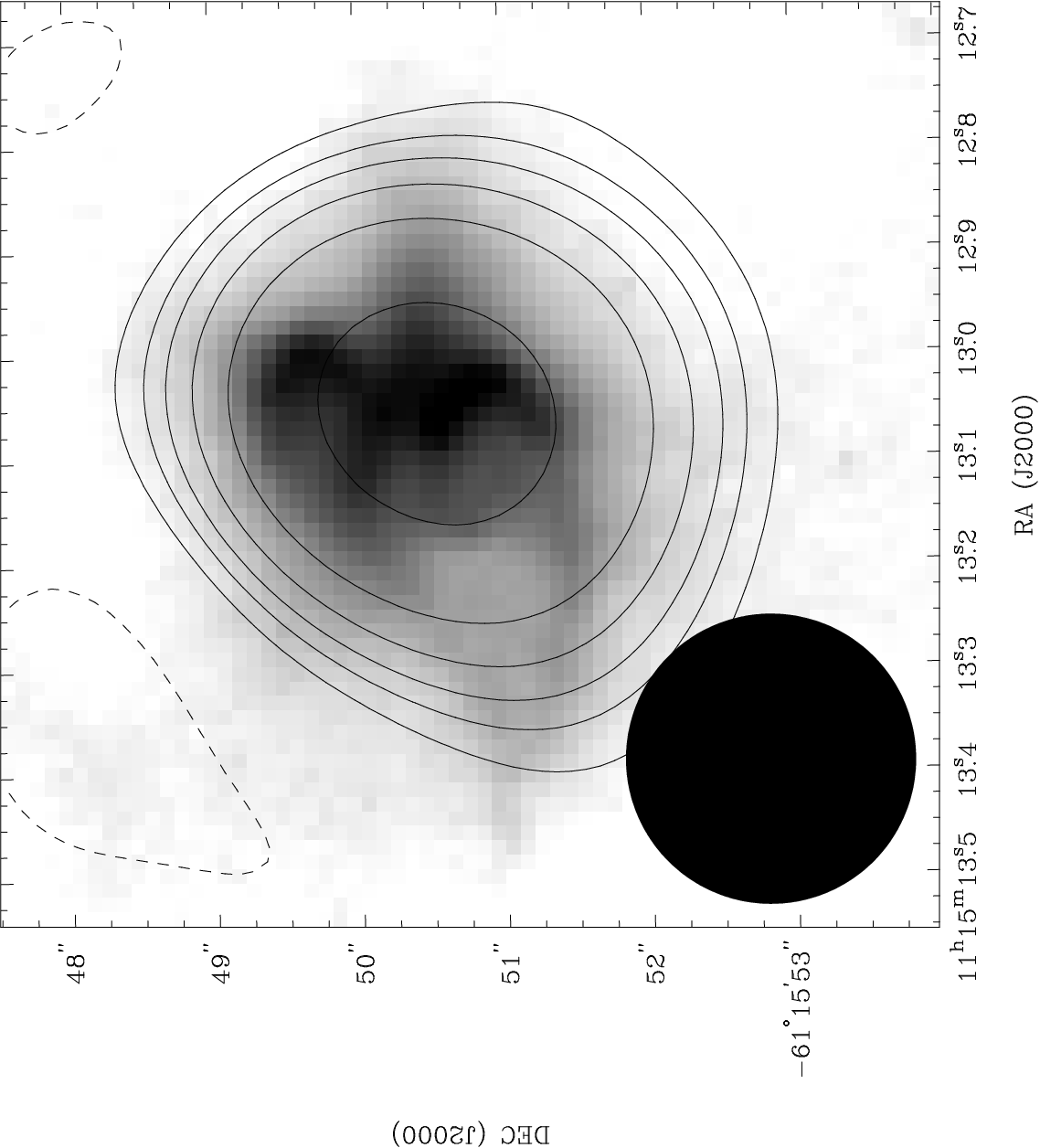}}
\rotatebox{-90}{\includegraphics[width=4.4cm]{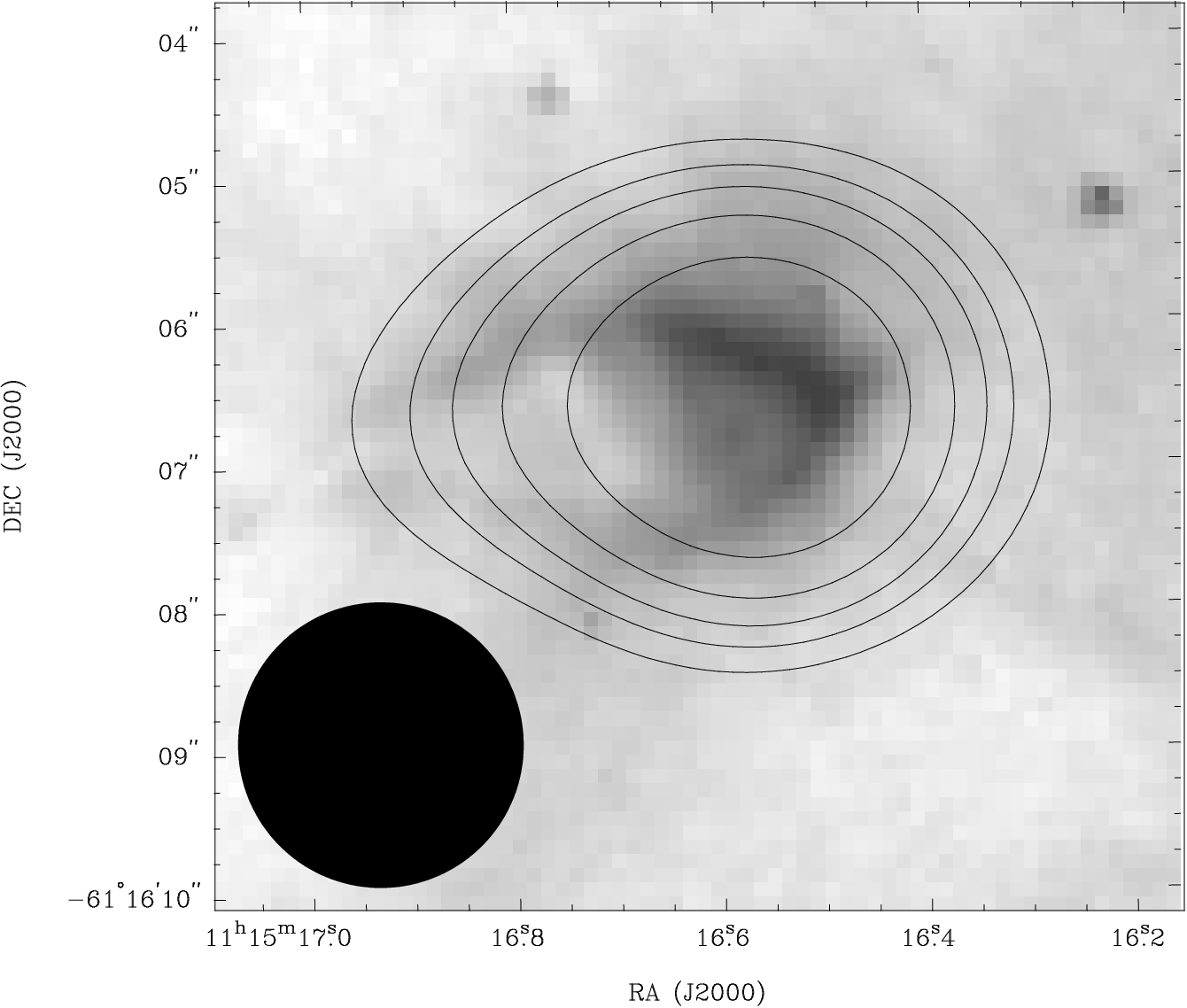}}
\rotatebox{-90}{\includegraphics[width=4.4cm]{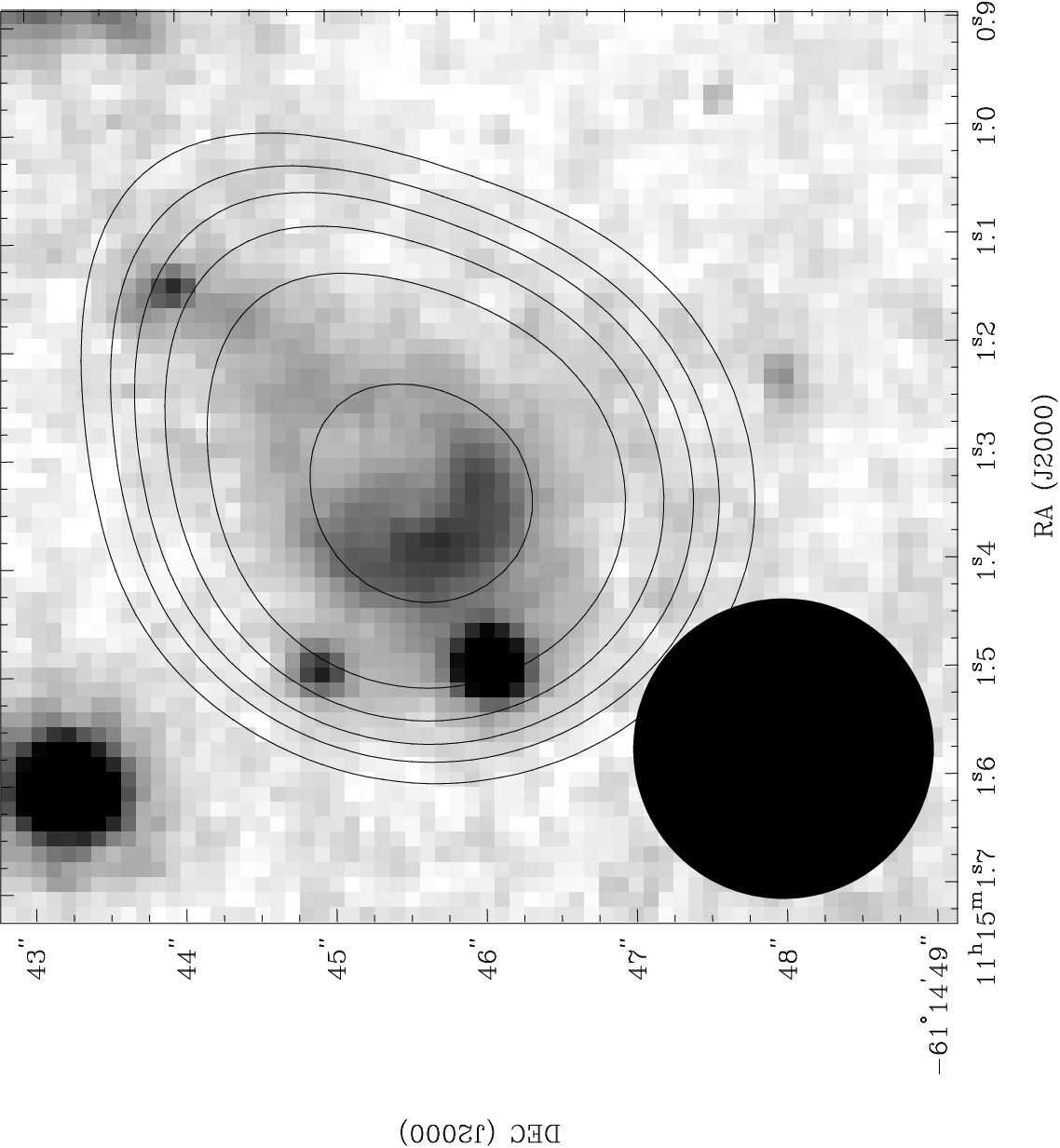}}
\end{tabular}
\caption{\ATCA\ 3~cm (upper row; $uv$ cutoff at 10 k$\lambda$) and 6-cm (lower row;
$uv$ cutoff at 25 k$\lambda$) radio continuum contour
maps of the three {\plike} sources, numbers 1--3 from left to right, overlaid
onto the \Ha\ image from \cite{brandner2} for P1 and P2, and onto the HST broad
band image for P3. The contour levels for the 3~cm maps are -0.45, 0.45, 0.75, 1.05, 1.5, 2.25
and 4.5 mJy\,beam$^{-1}$, and for the 6~cm maps are -0.9, 0.9, 1.5, 2.1, 3.0, 4.5
and 9.0 mJy\,beam$^{-1}$. The beam is shown in the bottom left
corner of each map. The \Ha\ images have been shifted 0\farcs5 to the North
and 0\farcs5 to the East in better alignment with the radio images. The arrows
indicate the directions projected onto the sky from which the stellar cluster
winds come. The cross near P3 indicates the position of the \Chandra\ X-ray
source.  \label{fig2}}
\end{figure*}

\clearpage

\begin{figure*} 
\rotatebox{-90}{\includegraphics[width=5.5cm]{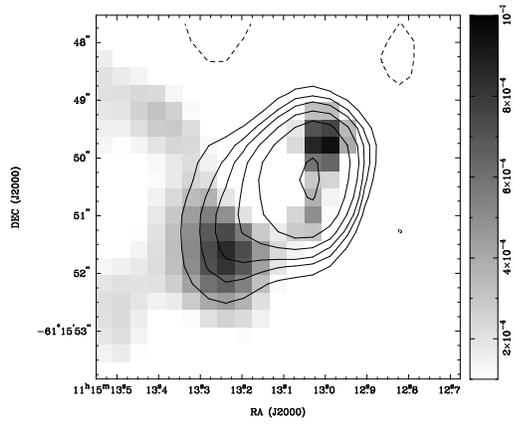}}
\caption{\ATCA\ 3~cm residual image of P1 after subtracting a single circular Gaussian
overlaid onto
its unsubtracted brightness contour map as shown in Fig.~2a. The contour levels correspond to
-0.3, 0.3, 0.5, 0.7, 1.0, 2.0 and 5.0 mJy\,beam$^{-1}$. P1 appears to possess two heads.
\label{fig3}}
\end{figure*}

\clearpage

\begin{figure*} 
\begin{tabular}{ll}
\rotatebox{-90}{\includegraphics[width=5.4cm]{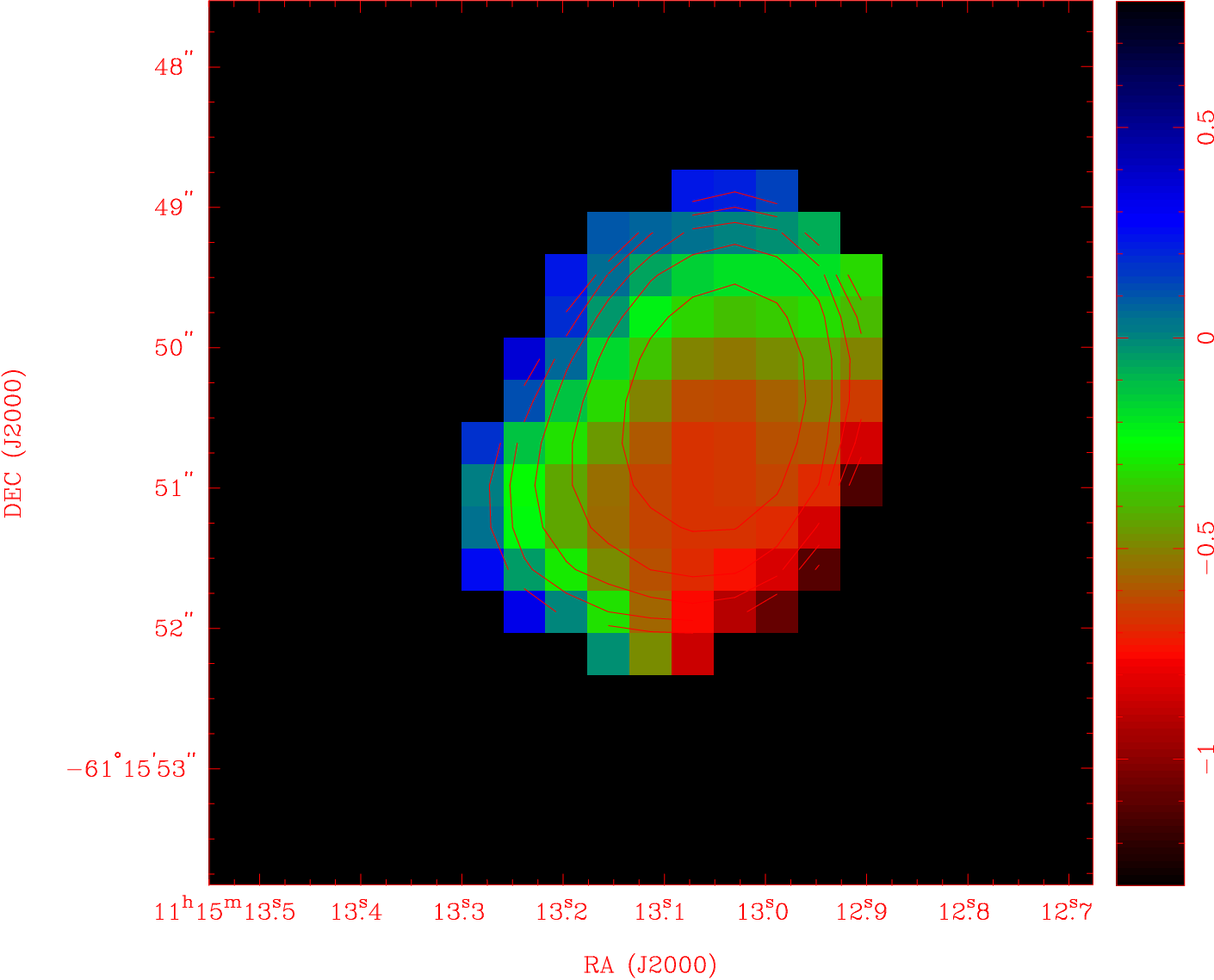}}
\rotatebox{-90}{\includegraphics[width=5.4cm]{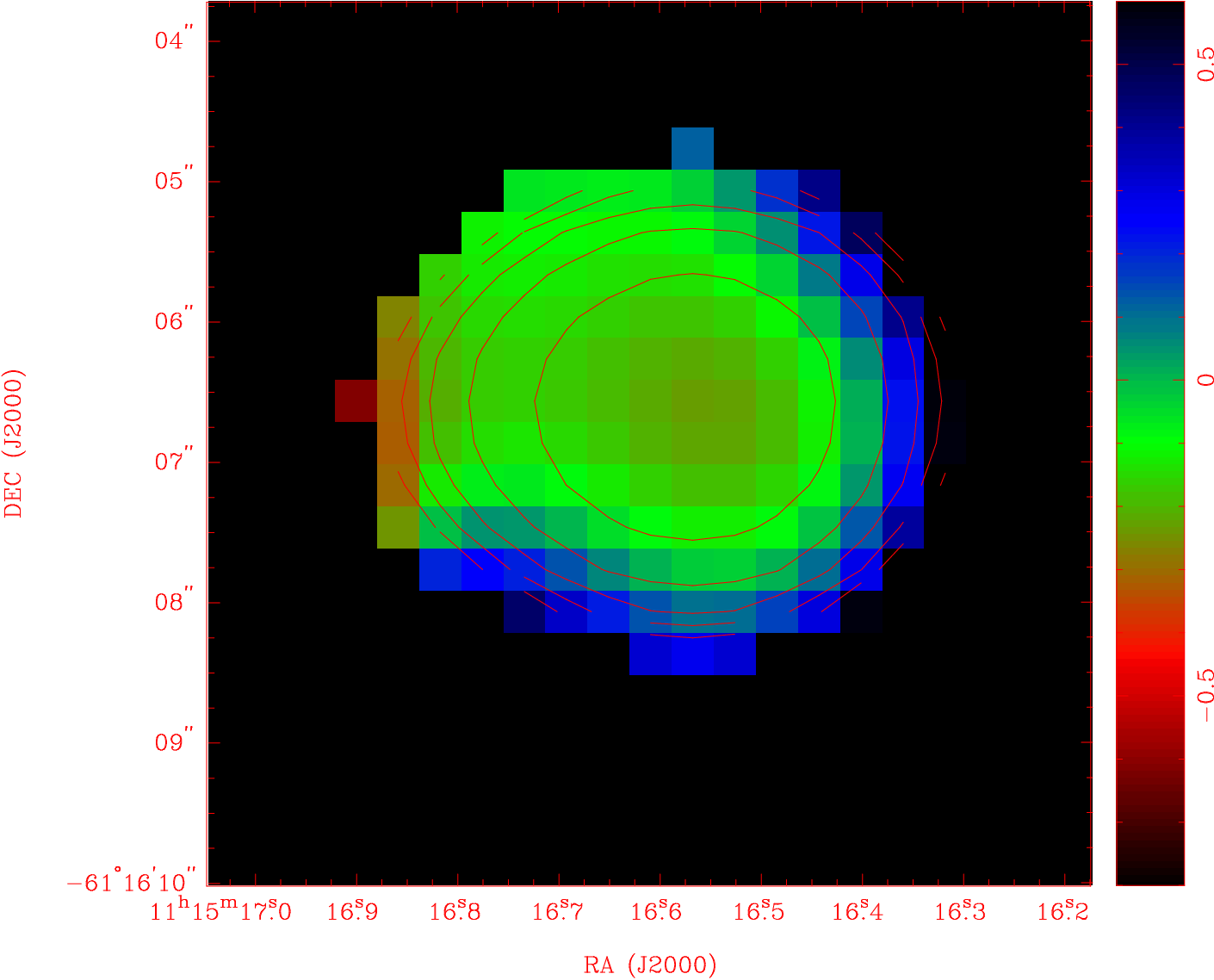}}
\rotatebox{-90}{\includegraphics[width=5.4cm]{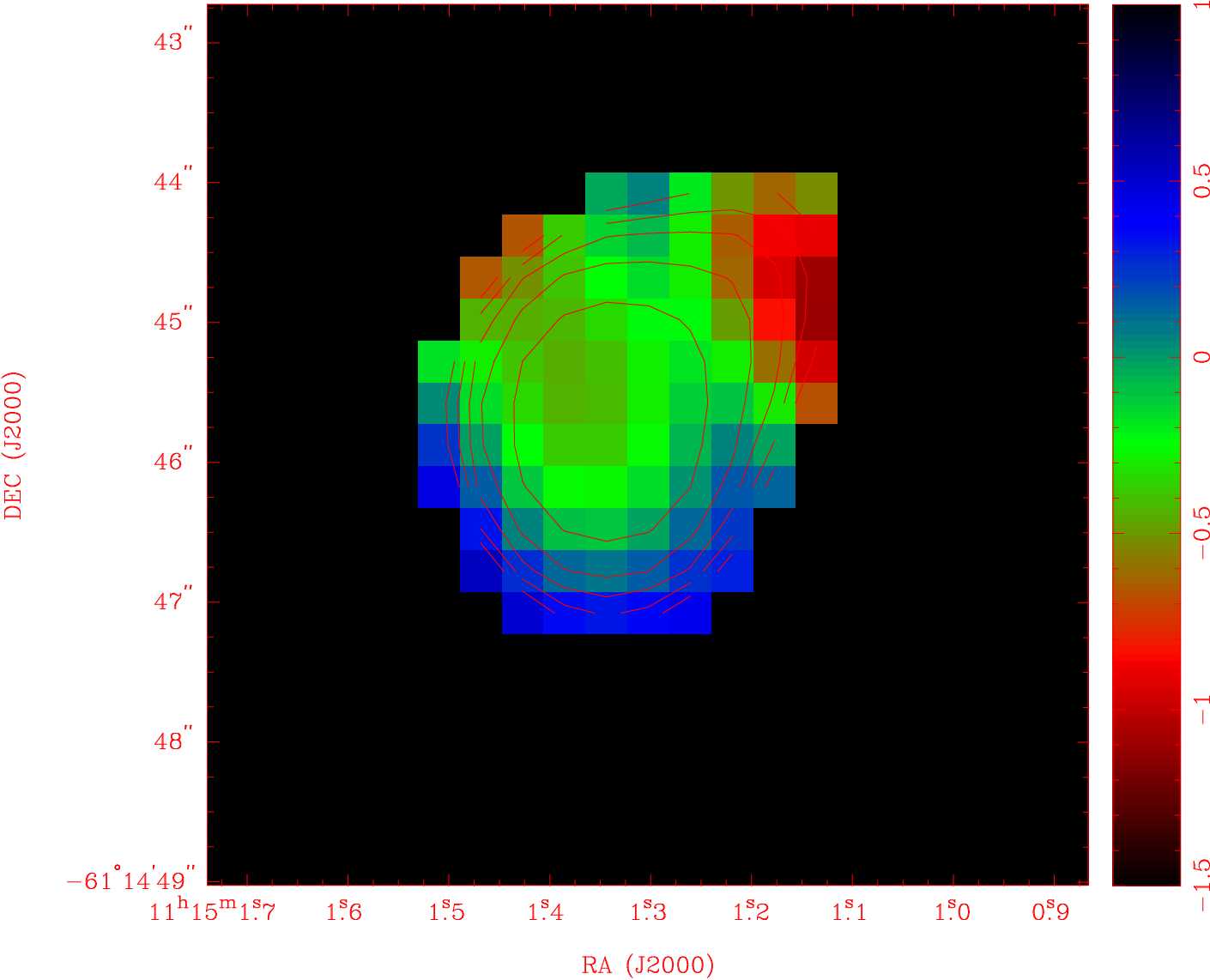}}
\end{tabular}
\caption{Examples of radio spectral index maps ($\alpha$; colour-coded) with the
corresponding uncertainty levels ($\Delta \alpha$; contours) for
P1, P2 and P3 (from left to right) derived from the radio maps shown in Fig.~2.
The contour levels are $\Delta
\alpha$ = 0.2, 0.3, 0.4, 0.5 and 0.6. The 3 and 6~cm flux densities used to
produce these maps have been clipped at 1.0 mJy\,beam$^{-1}$. Note the large
spectral index uncertainties at the rim of the {\proplyd}s due to the low
flux densities there. \label{fig4}}
\end{figure*}

\clearpage

\begin{table*} 
\caption{Observing parameters.\label{table1}}
\begin{tabular}{lccccc}
\tableline
Date (local time) & February 9/10 & April 8/9   & June 19  & Sept. 13/14 & Nov. 8/9 \\
Configurations    &  6A           & 6D          & 6B       & 6A           & 6C \\
Time on source    & 561 min.      & 514 min.    & 480 min. & 513 min.     & 543 min. \\
Pointing position & & &   \\
       RA (J2000) & \multicolumn{5}{c}{ 11\h\,15\m\,07\s }       \\
      DEC (J2000) & \multicolumn{5}{c}{ --61\degr\,16\arcmin\,00\arcsec} \\
Total bandwidth   & \multicolumn{5}{c}{ 128 MHz   }          \\
No. of channels   & \multicolumn{5}{c}{  32       }          \\
Frequencies       & \multicolumn{5}{c}{ 8640 MHz, 4800 MHz}          \\
Beam size      & \multicolumn{5}{c}{$\sim$1\arcsec $\times$ 1\arcsec (3~cm)} \\
                  & \multicolumn{5}{c}{$\sim$2\arcsec $\times$ 2\arcsec (6~cm)} \\
average r.m.s. near & \multicolumn{5}{c}{$\sim$0.1 (3~cm)} \\
ProPlyDs [mJy\,beam$^{-1}$]  & \multicolumn{5}{c}{$\sim$0.2 (6~cm)} \\
Flux calibrator   & \multicolumn{5}{c}{ 1934--63   }        \\
Phase calibrator  & \multicolumn{5}{c}{ 1059--63   }         \\
\tableline
\end{tabular}
\end{table*}

\clearpage

\begin{deluxetable}{lcccccc} 
\tablecaption{Position, fluxes and spectral indices of the {\proplyd}s.
 \label{table2}}
\tablecolumns{7}
\tabletypesize\footnotesize
\rotate
\tablewidth{17.5cm}
\tablehead{
\colhead{Name} & \multicolumn{2}{c}{P1} &
\multicolumn{2}{c}{P2} & \multicolumn{2}{c}{P3}\\
\colhead{$\lambda$} & \colhead{3~cm} & \colhead{6~cm} & \colhead{3~cm} &
\colhead{6~cm} & \colhead{3~cm} & \colhead{6~cm}}
\startdata
Radio position\tablenotemark{1}
               & 11:15:13.042, & 11:15:13.058,
               & 11:15:16.557, & 11:15:16.579,
               & 11:15:01.336, & 11:15:01.330, \\
RA,DEC (J2000) & --61:15:50.38 & --61:15:50.53
               & --61:16:06.57 & --61:16:06.62
               & --61:14:45.69 & --61:14:45.53\\
\hline
Peak flux & $4.3\pm 0.5$ & $11.5\pm 1.6$
          & $3.2\pm 0.4$ & $7.8\pm 1.1$
          & $5.6\pm 0.7$ & $10.9\pm 1.4$ \\
(mJy\,beam$^{-1}$) & & & & & &\\
\hline
Integrated & & & & & & \\
flux\tablenotemark{2} (mJy) & $13.9\pm 1.1$ & $18.1\pm 1.2$
                & $11.3\pm 1.4$ & $12.1\pm 0.7$
                & $14.1\pm 1.0$ & $16.5\pm 0.7$ \\
\hline
Spectral index& \multicolumn{2}{c}{$-0.5\pm 0.2$} &
                 \multicolumn{2}{c}{$-0.1\pm 0.2$} &
                 \multicolumn{2}{c}{$-0.3\pm 0.2$}\\
of whole source\tablenotemark{3} & & & & & & \\
\hline
\hline
$I_{\rm H\alpha}$\tablenotemark{4} (cm$^{-2}$\,s$^{-1}$)
             & 0.56 & 0.56 & 0.20 & 0.20 & \nodata & \nodata \\
\hline
Predicted & & & & & & \\
radio flux\tablenotemark{5} (mJy) & $1.56\pm 0.39$ & $1.65\pm 0.41$
               & $0.56\pm 0.14$ & $0.59\pm 0.15$ & \nodata & \nodata \\
\enddata
\tablecomments{The position of P3 given by Brandner et al. (2000; their
               Table~1) contains a typographical error.}
\tablenotetext{1}{Positions correspond to the radio locations of the 
                  {\proplyd}\,\,\,heads.}
\tablenotetext{2}{The quoted 3~cm and 6~cm integral fluxes are
                  derived from images with different resolution.}
\tablenotetext{3}{Flux density $S_\nu \propto \nu^{\alpha}$, $\alpha=$
                  spectral index.}
\tablenotetext{4}{\Ha\ flux corrected for an assumed foreground extinction
                  in \Ha\ of A$_{\rm H\alpha}$ = 4 mag.}
\tablenotetext{5}{See Brandner et al. (2000): assuming an electron temperature
                  for the Donised gas of $10^4$~K; the flux uncertainty is less
		  than 25\% due to attenuation by the neutral envelope of the
		  {\proplyd} itself as estimated by McCullough et al. 1995 
		  (see text).}
\end{deluxetable}

\clearpage

\begin{table} 
\caption{Fluxes and derived extinction\label{table3}}
\begin{tabular}{llccc}
\tableline
ProPlyD                    &                  &  1  &  2  &  3  \\
\tableline
3~cm peak flux             &[mJy\,beam$^{-1}$]& 4.3 & 3.2 & 5.6 \\
predicted \Ha\ peak flux&[cm$^{-2}$\,s$^{-1}$]& 1.5 & 1.1 & 2.0 \\
derived $A_{\rm H\alpha}$  & [mag]            & 5.1 & 5.7 & --- \\
\tableline
\end{tabular}
\end{table}

\clearpage

\begin{table*} 
\caption{Comparison with other known ProPlyDs\tablenotemark{(1)}.\label{table4}}
\begin{tabular}{llccc}
\tableline
              &         & Orion Nebula & Lagoon Nebula  &                 \\
              &         & M\,42        & M\,8, NGC~6523 & NGC~3603        \\
\tableline
distance      & [kpc]   & 0.45         & 1.8            & 6.1             \\
central star(s) &     &$\Theta^1$ Ori C& Herschel 36    & cluster         \\
star type(s)  &         &  O7          &    O7.5V       & 3 WNs+ $\sim$70
                                                            O stars       \\
$L_{\rm bol}$ & [\Lsun] & $\sim 10^5$  & $\sim 10^5$    & 2 $\times 10^7$ \\
Lyman flux    & [photons\,s$^{-1}$]    & $\sim 8\cdot 10^{48}$  
				       & 2 $\times 10^{48}$ & $10^{51}$   \\
extinction    & [mag]   &  4--6        &     $\sim 5$        & 4--6       \\ 
masers        &         &  H$_2$O, SiO & none known & OH, H$_2$O,
CH$_3$OH\\ \tableline
number of ProPlyDs  &   & $> 150$      & 1 = G5.97--1.7 &    3            \\
head size     & [AU]    & 45--355      & 1080           & 7200--10800     \\
distance to central stars & [pc] & 0.01--0.15 & 0.024   & 1.3, 2.2, 2.0   \\
radio flux    & [mJy]   &  0.3--1.4    & 17             & 10--12          \\   
brightness temp. & [K]  &  $\sim 10^2-10^3$ & $\sim 1500$    & 30--90          \\ 
emission measure & [pc\,cm$^{-6}$] 
              & $8\cdot 10^6$ & $(0.3-2)\cdot 10^8$ &$(3-9)\cdot 10^6f_{\rm{th}}^{-2/3}$\tablenotemark{(2)}\\
electron density & [cm$^{-3}$] & $10^5-10^6$ & $(4-20)\cdot 10^4$&$\sim 10^4$\\
mass loss rate   & [\Msun\,yr$^{-1}$]  & $1.2\cdot 10^{-7}$ 
				       & $7\cdot 10^{-7}$  & 10$^{-5}$     \\ 
bow shock distance &[AU]& $\sim 100$   & 540            & 44400,17400,\nodata\\
velocity      & [\kms]  & $\sim 10-15$ &  $\sim 10$     & 10--25    \\  
evap.  time scale&[years]& $\sim 10^4$ & $\sim 10^5$    & $\sim 10^5$   \\  
ProPlyD mass  & [\Msun] & 0.01--0.1    & $\sim 0.1$     & $\sim 1-10$   \\   
disk radius   & [AU] &  27--175        & 160            & 3400          \\   
disk mass     & [\Msun] & $\sim (0.5-20)\cdot 10^{-3}$ 
                                & $\sim 6\cdot 10^{-2}$ & not known     \\    
\tableline  
\end{tabular}

\tablenotetext{(1)}{Detailed information on the NGC~2024 ProPlyD is not
                  available.}
\tablenotetext{(2)}{See Sect. 3.4.}
\tablerefs{Stecklum et al. 1998, Brandner et al. 2000, Henney \& O'Dell 1999,
           McCullough et al. 1995} 
\end{table*}

\end{document}